\documentclass[a4paper,10pt]{article}
\pdfoutput=1 

\usepackage{jheppub} 

\usepackage[T1]{fontenc} 

\usepackage{amsfonts}
\usepackage{amsmath}
\usepackage{amsthm}
\usepackage{amssymb}
\usepackage{array}
\usepackage{dcolumn}
\usepackage{placeins}
\usepackage[svgnames]{xcolor}

\usepackage{graphics}
\usepackage{tikz}
\usepackage{graphicx}
\usetikzlibrary{positioning}
\usepackage{caption}
\usepackage{subcaption}
\usepackage{adjustbox}
\usepackage{gensymb}
\usepackage{breqn}
\usepackage{braket}
\usepackage{enumitem}
\usepackage{makecell}
\usepackage{ragged2e}
\usepackage{comment}
\usepackage{url}
\usepackage{placeins}
\usepackage{blindtext}

\usepackage{multirow}
\usepackage{bm}
\usepackage{float}

\newcommand{\be}{\begin{equation}}
\newcommand{\ee}{\end{equation}}
\newcommand{\bea}{\begin{eqnarray}}
\newcommand{\eea}{\end{eqnarray}}
\newcommand{\beq}{\begin{equation}}
\newcommand{\eeq}{\end{equation}}
\newcommand{\beqn}{\begin{eqnarray}}
\newcommand{\eeqn}{\end{eqnarray}}
\def\tev{\, {\rm TeV}}
\def\gev{\, {\rm GeV}}
\def\mev{\, {\rm MeV}}

\def\s{\, {\rm s}}

\newcommand{\gsim}{\lower.7ex\hbox{$\;\stackrel{\textstyle>}{\sim}\;$}}
\newcommand{\lsim}{\lower.7ex\hbox{$\;\stackrel{\textstyle<}{\sim}\;$}}
\newcommand{\fb}{\rm fb}

\preprint{APS/123-QED, MI-TH-771, KIAS-P22060}

\title{\boldmath Probing an MeV-Scale Scalar Boson in Association with a TeV-Scale Top-Quark Partner at the LHC}

\author[a]{Bhaskar Dutta,}
\author[a, b]{Sumit Ghosh,}
\author[c]{Alfredo Gurrola,}
\author[c]{Dale Julson,}
\author[a]{Teruki Kamon,}
\author[d]{and Jason Kumar}

\affiliation[a]{Mitchell Institute for Fundamental Physics and Astronomy, Department of Physics  and Astronomy, Texas A$\&$M University, College Station, Texas 77843,  USA}

\affiliation[b]{School of Physics, Korea Institute for Advanced Study, Seoul 02455, Korea}

\affiliation[c]{Department of Physics and Astronomy, Vanderbilt University, Nashville, TN, 37235, USA}

\affiliation[d]{Department of Physics and Astronomy, University of Hawaii, Honolulu, HI 96822, USA}

\emailAdd{dutta@physics.tamu.edu}
\emailAdd{ghosh@kias.re.kr}
\emailAdd{alfredo.gurrola@vanderbilt.edu}
\emailAdd{dale.a.julson@vanderbilt.edu}
\emailAdd{t-kamon@tamu.edu}
\emailAdd{jkumar@hawaii.edu}

\abstract{Searches for new low-mass matter and mediator particles have actively been pursued at fixed target experiments and at $e^+e^-$ colliders. It is challenging at the CERN LHC, but they have been searched for in Higgs boson decays and in $B$ meson decays by the ATLAS and CMS Collaborations, as well as in a low transverse momentum phenomena from forward scattering processes (e.g., FASER). We propose a search for a new scalar particle in association with a heavy vector-like quark. We consider the scenario in which the top quark ($t$) couples to a light scalar $\phi’$ and a heavy vector-like top quark $T$. We examine single and pair production of $T$ in $pp$ collisions, resulting in a final state with a top quark that decays purely hadronically, a $T$ which decays semileptonically ($T$ $\rightarrow$ $W$ + $b$ $\rightarrow$ $\ell$ $\nu$ $b$), and a $\phi'$ that is very boosted and decays to a pair of collimated photons which can be identified as a merged photon system. The proposed search is expected to achieve a discovery reach with signal significance greater than 5$\sigma$ (3$\sigma$) for $m(T)$ as large as 1.8 (2) TeV and $m(\phi')$ as small as 1 MeV, assuming an integrated luminosity of 3000 fb$^{-1}$. This search can expand the reach of $T$, and demonstrates that the LHC can probe low-mass, MeV-scale particles.
}


\begin{document} 
\maketitle
\flushbottom

\section{Introduction}
\label{sec:intro}

Recently, there has been great interest in models of beyond-the-Standard Model (BSM) physics involving new low-mass matter and mediator particles.  A promising  method for testing these models is the use of beam experiments, where the focus is typically on processes with relatively low momentum transfer, in which the low-mass mediators are produced at high longitudinal boost (see, for example,~\cite{Batell:2017kty,Bauer:2018onh,Abdullah:2018ykz}).  Indeed, these types of processes are the typical production mechanisms for mediators which arise in the low-energy effective field theory.  The visible decays of long-lived mediators can then be observed at displaced detectors (e.g., FASER~\cite{FASER:2019aik}), with theintervening shielding serving to remove the background arising from the beam itself.

The Large Hadron Collider (LHC), however, excels at producing new heavy particles in high momentum transfer processes. Thus, current LHC searches for light mediators $a$ (e.g., axion-like particles) are carried out in the context of SM Higgs decays ($H \to aa(Za)$), $B$ meson decays ($B \to Xa$), or in proton-proton ($pp$) production mechanisms resulting in associated top quarks. One strategy often pursued in those analyses is a dilepton selection targeting the $a \to \ell\ell$ decay mode. However, the model parameter space (mass, lifetime, and coupling) that can be probed by those searches is often limited, especially for an MeV-scale mediator, where $a \to \ell\ell$ is not kinematically allowed. The sensitivity to new physics models with low mass mediators may be further suppressed by the low momentum attributed to the $a$ decay products, making it difficult to suppress SM backgrounds and experimentally trigger on interesting events.

Generically, if BSM physics involves new low-mass particles and non-renormalizable opertors, then a UV completion of the low-energy model can introduce couplings between the new low-mass particles and new high-mass particles.  This provides for an interesting opportunity, in which one searches for event topologies with both new low-mass and high-mass particles.

We devise a LHC search strategy for a light mediator particle produced in association with a TeV-scale partner particle of the top quark ($t$). The heavy top-partner particle can be copiously produced at the LHC because it couples to the SM quarks and gluons, and its highly energetic decay products can be observed in association with a mediator particle containing substantial transverse momentum. If the mediator decays promptly into the SM particles, then the resulting SM particles will be detected in the central region of the detector and will be energetic enough to significantly reduce SM backgrounds. As a result of this search strategy, the LHC can expand the reach of heavy top-partners and may discover even MeV-scale mediators, which are otherwise difficult to probe at a hadron collider using traditional search strategies. There exist a few search strategies for the heavy top partner in the literature~\cite{Cacciapaglia:2019zmj, Benbrik:2019zdp, Aguilar-Saavedra:2017giu, Dermisek:2019vkc, Chala:2017xgc}. In these scenarios, the heavy top partner  decays to top quark along with a heavy new scalar. However, they don’t address the case of a MeV-scale scalar.  

In this paper,the heavy fermion top-partner is denoted by $T$, which mixes with the top flavor eigenstate.  As a result, the flavor sector of the theory can be more complicated, and can introduce a coupling of $T$ to a new low-mass scalar mediator, denoted by $\phi'$.  In addition to the expected $T$ decays to $bW$, $tZ$ and $tH$, one can generically have a decay to $t\phi'$.  The $T-t-\phi'$ coupling thus allows processes, such as associated production, in which the final state includes $Tt\phi'$, with $\phi'$ decaying to $\gamma \gamma$.  We will see that the combination of $T$ and $t$ decay products, plus energetic electromagnetic activity, will provide for search sensitivity which exceeds that of searches for either $T$ or $\phi'$ in isolation. 

The scenario of new heavy quarks and light mediators leads to a variety of interesting phenomenological features.  For example, since the new heavy quarks can mix with Standard Model quarks, new contributions to quark flavor violation can arise.  The coupling of the light mediators to leptons can lead to lepton non-universality. 
Another class of model where a light scalar mediator can arise is that of the extended scalar sector model where the scalar sector of the SM is extended by new scalar fields. The mixing of the new scalar fields with the SM scalar doublet can give rise to a light scalar ($\lesssim \mathcal{O}(100)$~MeV) after the spontaneous breakdown of electroweak symmetry. For example, a physical scalar spectrum with one light scalar can be generated by extending the SM scalar sector by an additional doublet and one singlet scalar field~\cite{Dutta:2020scq}. Another example can be an extension by an additional doublet, one triplet and one singlet scalar field~\cite{Dutta:2022fdt}. Through the scalar mixing, the light scalar field couples to top quark. If we assume that this extended scalar sector is part of some underlying heavy physics where heavy vector-like fermions are present that can act as the heavy top-partner, we can generate a $T-t-\phi^\prime$ coupling. In these models $\phi^\prime$ can decay to $\gamma \gamma$ via a fermion loop.   

It should be noted that our interest is to develop a signature-based search strategy, not to focus on the details of any particular UV-complete model which might generate this signal.However, we develop a benchmark model in the appendix, in order to  have  concrete, physically realizable values for the relevant couplings and branching fractions.

We consider a new gauge group, $U(1)_{T3R}$~\cite{Pati:1974yy,Mohapatra:1974gc,Senjanovic:1975rk} with a sub-GeV dark Higgs boson~\cite{Dutta:2019fxn,Dutta:2020enk,Dutta:2021afo}).  Since only right-handed SM fermions (including the top quark) are charged under $U(1)_{T3R}$, the effective Yukawa coupling needs a Higgs and dark Higgs insertion, and is thus a non-renormalizable operator.  This scenario can be UV completed with the universal seesaw mechanism~\cite{Berezhiani:1983hm,Chang:1986bp,Davidson:1987mh,Rajpoot:1987fca,Babu:1988mw,Babu:1989rb}, yielding a fermionic partner for the top.

The new heavy quarks of this model can be produced at the LHC, and their decay products observed in association with an energetic $\phi^\prime$ in the central regions of the CMS~\cite{CMS:2008xjf} and ATLAS~\cite{ATLAS:2008xda} detectors. The decay $\phi^\prime \rightarrow \gamma \gamma$ will then produce a pair of collimated photons.  The large SM background can thus be reduced by requiring the photons to be energetic, and by tagging $t$ and bottom quarks ($b$) which also necessarily appear.

The rest of this paper is structured as follows. Section II provides an overview of current results from searches for vector-like fermions at the LHC. Section III provides details of how the Monte Carlo samples were produced for this study. In Section IV we discuss event selection criteria, and in Section V, main results.  We end with a short discussion in Section VI.

\section{Review of vector-like fermion searches} \label{sec:reviewofvectorlikefermionsearches}

\begin{figure}[tbp]
\subfloat[QCD Production]{\label{fig:sfig1}%
  \includegraphics[width=0.5\linewidth]{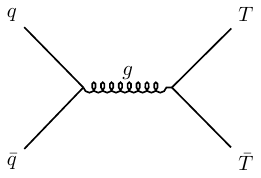}
}
\subfloat[Electroweak Production]{\label{fig:sfig2}%
  \includegraphics[width=0.5\linewidth]{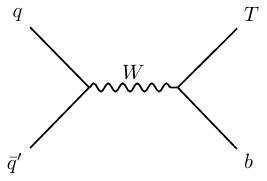}
}
\caption{Feynman diagrams for $T$ pair production via gluon-mediated QCD (left) and $T$ production via electroweak processes (right). These processes have been studied in previous experimental analyses.}
\label{fig:production_diagrams}
\end{figure}

Previous LHC searches for $T$ have been performed by ATLAS and CMS using $pp$ collisions at $\sqrt{s} = 8$ and $13$ TeV [10–24]. Those results focused on pair production via gluon mediated QCD processes from $q\bar{q}$ annihilation (Fig.~\ref{fig:sfig1}) or single $T$ production from electroweak processes containing associated quarks (Fig.~\ref{fig:sfig2}), followed by the $T$ decay to $bW$, $tZ$, or $tH$. The analyses which targeted the $bW$ decay mode utilized jet substructure observables in order to identify hadronic decays of boosted $W$ bosons, which allows for the reconstruction of the $T$ candidate. The analyses targeting the $tZ$ and $tH$ decay modes used a multiclassification boosted event shape algorithm to identify jets originating from boosted $t$, $Z$, and $H$. The $T$ branching fractions depend on the particular choice of model, and therefore the ATLAS and CMS searches are performed for various possible combinations of the $T$ branching fractions, resulting in 95\% confidence level upper limits on their masses ranging from 740 GeV to about 1.3 TeV. We use the notation $m(T)$ for the mass of the heavy quark eigenstate, to avoid confusion with the transverse mass kinematic variable, which we will denote as usual by $m_T$. 

In the case of $T$ pair production via QCD processes, the cross sections are known and only depend on the mass of the vector-like quark. Assuming a narrow $T$ decay width $(\Gamma/m(T) < 0.05$ or $0.1)$ and a $100$\% branching fraction to $bW$, $tZ$, or $tH$, those searches have placed stringent bounds on $m(T)$, excluding masses below about $1.2$ TeV at 95\% confidence level (CL). On the other hand, the excluded mass range is less stringent for $\{\mathrm{Br}(T\to tZ),\mathrm{Br}(T\to bW),\mathrm{Br}(T\to tH)\} = \{0.2,0.6,0.2\}$, excluding masses below about $0.95$ TeV. However, if the $T \rightarrow \phi' t$ decay is allowed, those limits must be re-evaluated. In particular, the authors of Ref.~\cite{Cacciapaglia:2019zmj} point out that the bounds on $m(T)$ can be about 500 GeV when $T \to t\phi'$ decays are allowed and/or when the $Br(T\to tH/bW)$ branching fractions are lower. Therefore, to allow for a broad and generic study, we consider benchmark scenarios in our toy model down to $m(T) = 500$ GeV.

 \begin{figure}[tbp]
 \begin{center} 
 \includegraphics[width=0.5\textwidth]{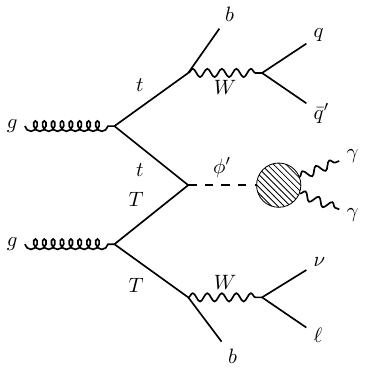}
 \end{center}
 \caption{``$T$-$t$ Fusion'' Feynman diagram.}
 \label{fig:FeynmanDiagram}
 \end{figure}

\section{Samples and simulation} \label{sec:samplesandsimulations}

In keeping with our notation for the new particle $T$, we will henceforth use $m(\phi')$ to denote the mass of the light scalar eigenstate.Given that this study is interested in scenarios where $m(\phi')$ is small, it is experimentally necessary to target an interaction topology which can provide ample boost to the $\phi'$ to aid in the reconstruction of its decay products. For this reason, the $T$-$t$ fusion interaction, as shown in Fig.~\ref{fig:FeynmanDiagram}, was studied due to its sufficiently large cross section over a wide range of $T$ masses, its unique topology that allows for sufficient background suppression, and the large boost given to the $\phi'$ and subsequent decay products via the more massive $T$ and $t$. In particular, we note that the $T$-$t$ fusion cross section is somewhat larger than the $T$ pair production cross section.  This is to be expected, as we are considering the parameter space for which $m(T) \gg m_t$. 

The $T$-$t$ fusion mechanism occurs when the protons at the LHC approach the collision point, and the gluons that make up each proton split their energy into a $t\bar{t}$ pair and a $T\bar{T}$ pair. One top quark and one heavy vector-like fermion interact with each other to produce a $\phi'$, while the other $t$ and $T$ create Standard Model decay products in the detector. We target events in which the top quark decays purely hadronically into a bottom quark and dijet pair ($t \to b W \to b q \bar{q}'$), the $T$ decays semileptonically into a $b$ quark, lepton, and neutrino, and the $\phi'$ produces two photons. Since the mass of the $t$-$T$ system which produces the $\phi'$ is much larger than $m(\phi')$ for the regions of parameter space studied in this paper, the large rest energy is converted into large momentum values for the $\phi'$.  

We assume that the heavy  fermionic top partner $T$ has the same quantum numbers under the unbroken symmetries of the Standard Model at the top quark, and that the low-mass scalar mediator $\phi'$ is neutral under those unbroken symmetries.  As a result, the top flavor eigenstate is generally a linear combination of the mass eigenstates $t$ and $T$.  We generically find that $T$ can couple to $bW$, $tZ$, $tH$ and $t \phi'$.  In the appendix, we develop a particular model in which the structure of these couplings are controlled by the symmetry group $U(1)_{T3R}$.  Choosing the new dimensionless couplings to take ${\cal O}(1)$ values, as described therein, we can obtain the rates for $T$ to decay to the various available final states, as a function of the $m(T)$.  Table~\ref{tab:Width} lists the relevant $T$ decay widths for a benchmark scenario with $m(T) = 2$ TeV. The $T$ decay width is relatively small compared to the mass of $T$ ($\frac{\Gamma_{T}}{m(T)} < 1$\%), and thus the ability to experimentally reconstruct the mass of $T$ is dominated by experimental/detector resolution. A more comprehensive discussion is provided in the appendix. As a benchmark for our analysis, we will take the branching fraction $\mathrm{Br}(T \rightarrow bW)$ to be $50\%$.

Note that the signal topology we consider can also be produced by events in which $T$ decays to  $tZ$, with $t$ decaying semileptonically and with $Z$ decaying invisibly.  Since the relative branching fraction between the $bW$ and $tZ$ channels is  model-dependent, we will preserve the generality of this study by considering a conservative analysis in which signal events from the $tZ$ channel are ignored.  Since the branching fraction for $Z \rightarrow \nu \nu$ is small, this conservative analysis is nevertheless a good approximation.

\begin{table}[tbp]
\centering
\begin{tabular}{c|l}
\hline
Process & Decay Width [GeV]  \\

&\\

\hline 
$T \to bW$ & ~~~~~~~~1.225 \\
$T \to tH$ & ~~~~~~~~0.0874 \\
$T \to tZ$ & ~~~~~~~~1.191 \\
$T \to t\phi'$ & ~~~~~~~~0.0881 \\
\hline
\end{tabular}
\captionsetup{justification   = RaggedRight,
             labelfont = bf}
\caption{\label{tab:Width} $T$ decay width for a benchmark scenario with 
$m(T) = 2$ TeV.
We assume that the $b$-quark has no mixing with new heavy fermions.  This 
scenario is discussed in more detail in the appendix.}
\end{table}

\begin{figure}[tbp]

 \begin{center} 
 \includegraphics[width=0.65\textwidth]{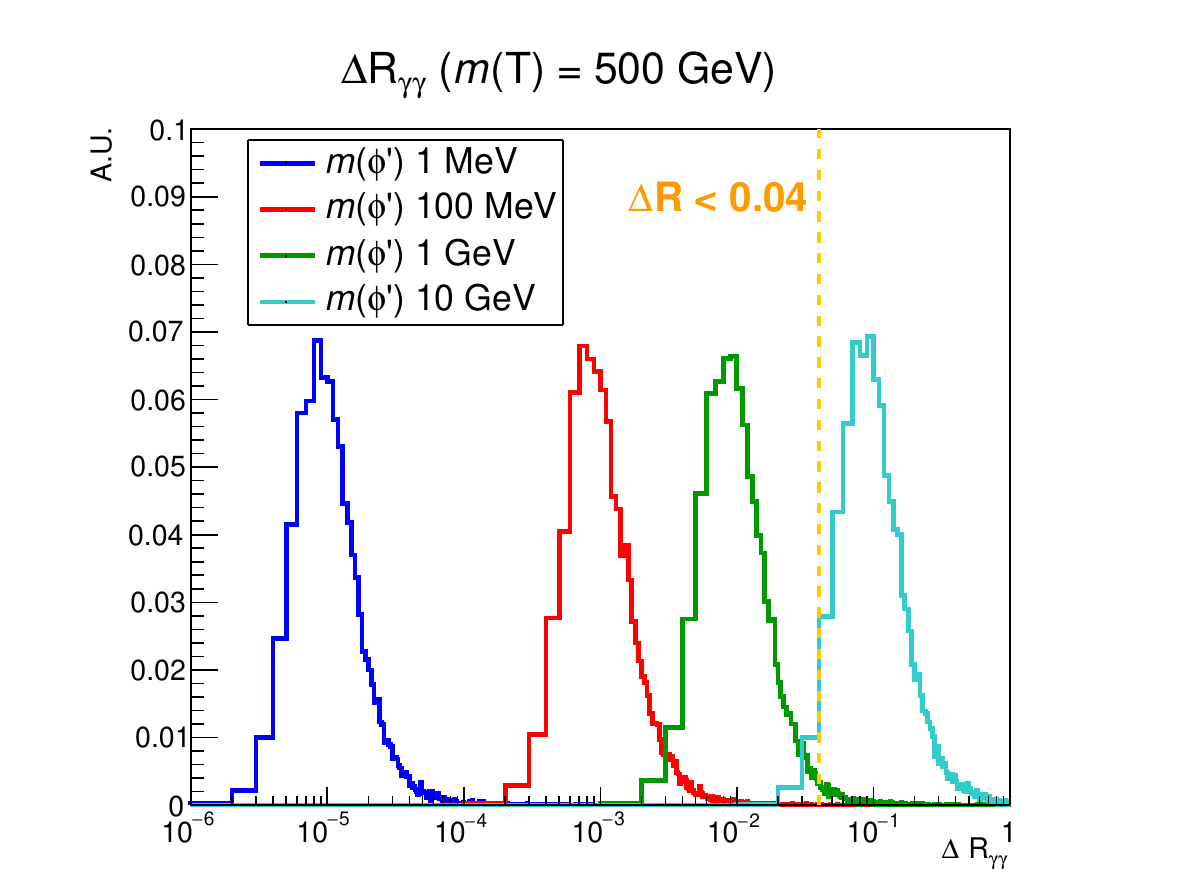}
 \end{center}
 \caption{The distribution of $\Delta R_{\gamma\gamma}$ for $m(\phi') = 1~\mev$ (blue), 
 $100~\mev$ (red), $1~\gev$ (green) and $10~\gev$ (cyan), assuming $m(T) = 500~\gev$.  The normalization of the $y$-axis  is arbitrary.  The vertical dashed orange line denotes $\Delta R_{\gamma\gamma} = 0.04$; for events to the left of this line, the two photons cannot be discriminated.} 
 \label{fig:photon_DeltaR}
 \end{figure}

Given the large boost of the $\phi'$, the $\phi'\to\gamma\gamma$ decay results in two collimated photons which are almost always reconstructed as a single merged photon system, depending on the spatial resolution of the electromagnetic calorimeter (ECAL). For example, the CMS ECAL is a homogeneous and hermetic calorimeter containing lead tungstate ($\textrm{PbWO}_{4}$) scintillating crystals with front face cross sections of around 2.2 cm $\times$ 2.2 cm, resulting in a diphoton spatial resolution of $\Delta R_{\gamma\gamma} =\sqrt{(\Delta\phi_{\gamma\gamma})^2+(\Delta\eta_{\gamma\gamma})^2} \sim 0.04$, where $\Delta\phi_{\gamma\gamma}$ is the azimuthal angle between the two photons, and $\Delta\eta_{\gamma\gamma}$ is the pseudorapidity gap between the two photons. As shown in Fig.~\ref{fig:photon_DeltaR}, for any $m(\phi') < 1$ GeV, the two photons produced from the $\phi'$ decay will almost always have $\Delta R<0.04$. These diphoton pairs will therefore be reconstructed as a single photon in the detector. These results are consistent with those found in Ref.~\cite{Sheff:2020jyw}. We note there is an effort to reconstruct two photons from a light scalar using machine learning technique in the LHC experiments~\cite{CMS:2022wjj, CMS:2022fyt}.  In order to emulate this merged photon effect, any diphoton pairs in both the signal and background samples with $\Delta R<0.04$ were clustered into a single photon object in the detector simulation software (described later). It is noted that Ref.~\cite{Knapen:2021elo} have studied boosted diphoton systems from decays of light mediators, and have concluded that an analysis strategy utilizing isolated diphoton triggers, as is traditionally done in $X\to \gamma \gamma$ resonant searches at CMS and ATLAS, does not work. Therefore, in these studies we highlight the importance of the unique topology, which contains boosted $t$/$W$’s, to aid in establishing a suitable experimental trigger, without biasing the merged photon system, and thus maintaining high signal acceptance to achieve discovery potential.

A similar effect is observed for the energetic $t$ produced through the $T$-$t$ fusion interaction, in which a portion of the $t$ decay products cannot be fully reconstructed independently of each other. This results in three different scenarios of $t$ reconstruction: a fully merged scenario, in which the decay products of the $W$ boson and the $b$ quark are so collimated that they are reconstructed as a single ``fat jet’’ (henceforth referred to as a FatJet, FJ); a partially merged scenario, in which the decay products of the $W$ boson form a single FatJet but the $b$ quark can still be separately identified; and an un-merged scenario, in which all decay products can be independently identified. Each scenario has an associated tagging efficiency and misidentification rate, which depends on the boosted $t$/$W$ algorithm. Following Refs.~\cite{CMS:2020poo,ATLAS:2018wis}, we consider two possible ``working points'' for the identification of the fully merged top decays: ($i$) a ``Loose'' working point with 85\% top tagging efficiency and 11-25\% misidentification rate, depending on the FatJet transverse momentum ($p_{T}^{FJ}$); and ($ii$) a ``Tight'' working point with 50\% top tagging efficiency and 4-10\% misidentification rate, depending on $p_{T}^{FJ}$. Similarly, for the partially merged scenario we consider a ``Loose'' working point with 50\% $W$ tagging efficiency and 1-2\% misidentification rate, and a ``Tight'' working point with 25\% $W$ tagging efficiency and 0.2-0.5\% misidentification rate. The choice of boosted $t$/$W$ working points is determined through an optimization process which maximizes discovery reach, as discussed in the next section. The ``Loose'' working  points for both the fully merged $t$ and partially merged $W$ were ultimately shown to provide the best sensitivity and therefore chosen for this study. For now we note that the contribution from SM backgrounds with a misidentified boosted $t$/$W$ is negligible, and thus our discovery projections are not sensitive to uncertainties related to the boosted $t$/$W$ misidentification rates.

Due to the above considerations, the ultimate final-state of interest will consist of: a single highly energetic photon, a (possibly boosted) top tagged system, a $b$ quark, a lepton, and large missing transverse momentum ($\vec{p}_{T}^{\textrm{~miss}}$). For the partially merged and un-merged scenarios, there will be two $b$ quarks present in the final state (one of which is part of the top tagged system). It is important to be able to differentiate between the $b$ quark associated with the hadronic decay of the top quark and the $b$ quark associated with the $T$ decay, for instance when attempting to reconstruct/measure the mass of the $T$ using its decay products. An identification scheme was therefore employed in which the $b$ quark located closest in $\Delta R$ to the dijet pair (from $W\to jj$) was labeled the ``hadronic decay'' $b$ quark ($b_{h}$), while the further $b$ quark would be referred to as the ``leptonic decay'' $b$ quark ($b_{\ell}$). This scheme mimics what is possible experimentally, where one cannot identify the exact decay origins of any particular particle. It was found that this scheme correctly labeled each $b$ quark with an accuracy of greater than 80\%, depending on $m(T)$.

Sample events for $pp$ collisions at $\sqrt{s}=13$ TeV for both signal and background were generated using MadGraph5\_aMC (v2.7.3)~\cite{Alwall:2014hca}. Hadronization was performed with PYTHIA (v8.2.05) \cite{Sjostrand:2014zea}. Detector effects were included through Delphes (v3.4.1) \cite{deFavereau:2013fsa}, using the CMS input card with 140 average pileup interactions. Signal samples were created for $m(T)$ ranging from 500 GeV to 2500 GeV. The large mass difference between $T$ and $\phi'$ resulted in kinematic distributions for $\phi'$ that were similar regardless of $m(\phi')$ mass (for masses up to 10 GeV).

\begin{table}[tbp]
\centering
\subfloat[150 fb$^{-1}$ integrated luminosity.]{
\label{tab:lifetime_150}%
\renewcommand{\arraystretch}{1.25}
\begin{tabular}{|l|c|c|c|c|c|}
\cline{3-6}
\multicolumn{2}{c|}{} & \multicolumn{4}{c|}{$\bm{m}(\phi')$ $\textbf{[MeV]}$} \\
\cline{3-6}
\multicolumn{2}{c|}{} & 0.1 & 1 & 10 & 100 \\
\cline{1-6}
\multirow{7}{*}{\rotatebox[origin=c]{90}{$\bm{m(T)}$ $\mathbf{[GeV]}$}} & 500  & 2.6$\times10^{-15}$ & 2.6$\times10^{-14}$ & 2.6$\times10^{-13}$ & 2.6$\times10^{-12}$ \\
\cline{2-6}
& 750  & 2.4$\times10^{-15}$ & 2.4$\times10^{-14}$ & 2.4$\times10^{-13}$ & 2.4$\times10^{-12}$ \\
\cline{2-6}
& 1000 & 1.4$\times10^{-15}$ & 1.4$\times10^{-14}$ & 1.4$\times10^{-13}$ & 1.4$\times10^{-12}$ \\
\cline{2-6}
& 1250 & 5.6$\times10^{-16}$ & 5.6$\times10^{-15}$ & 5.6$\times10^{-14}$ & 5.6$\times10^{-13}$ \\
\cline{2-6}
& 1500 & N/A & N/A & N/A & N/A \\
\cline{2-6}
& 1750 & N/A & N/A & N/A & N/A \\
\cline{2-6}
& 2000 & N/A & N/A & N/A & N/A \\ \hline
\end{tabular}
}
\par
\subfloat[3000 fb$^{-1}$ integrated luminosity.]{
\label{tab:lifetime_3000}%
\renewcommand{\arraystretch}{1.25}
\begin{tabular}{|l|c|c|c|c|c|}
\cline{3-6}
\multicolumn{2}{c|}{} & \multicolumn{4}{c|}{$\bm{m}(\phi')$ $\mathbf{[MeV]}$} \\
\cline{3-6}
\multicolumn{2}{c|}{} & 0.1 & 1 & 10 & 100 \\
\cline{1-6}
\multirow{7}{*}{\rotatebox[origin=c]{90}{$\bm{m(T)}$ $\mathbf{[GeV]}$}} & 500 & 2.7$\times10^{-15}$ & 2.7$\times10^{-14}$ & 2.7$\times10^{-13}$ & 2.7$\times10^{-12}$ \\
\cline{2-6}
& 750  & 2.6$\times10^{-15}$ & 2.6$\times10^{-14}$ & 2.6$\times10^{-13}$ & 2.6$\times10^{-12}$ \\
\cline{2-6}
& 1000 & 2.3$\times10^{-15}$ & 2.3$\times10^{-14}$ & 2.3$\times10^{-13}$ & 2.3$\times10^{-12}$ \\
\cline{2-6}
& 1250 & 1.5$\times10^{-15}$ & 1.5$\times10^{-14}$ & 1.5$\times10^{-13}$ & 1.5$\times10^{-12}$ \\
\cline{2-6}
& 1500 & 9.0$\times10^{-16}$ & 9.0$\times10^{-15}$ & 9.0$\times10^{-14}$ & 9.0$\times10^{-13}$ \\
\cline{2-6}
& 1750 & 4.8$\times10^{-16}$ & 4.8$\times10^{-15}$ & 4.8$\times10^{-14}$ & 4.8$\times10^{-13}$ \\
\cline{2-6}
& 2000 & N/A & N/A & N/A & N/A \\ \hline
\end{tabular}
}
\label{tab:lifetime}\caption{Longest lifetime of $\phi'$ [s] for which 5$\sigma$ discovery can be achieved. ``N/A'' indicates no 5$\sigma$ discovery is possible (due to small cross sections).}
\end{table}

Total event yields are parameterized using $N=\sigma\times\mathcal{L}\times\epsilon$, where $N$ represents the total number of events, $\mathcal{L}$ is the integrated luminosity scenario being considered (for this study, 150 fb$^{-1}$ and 3000 fb$^{-1}$), and $\epsilon$ represents any efficiencies which might reduce the total event yield (reconstruction efficiencies, etc.). All production cross sections were computed at tree-level.  As the $k$-factors associated with higher-order corrections to QCD production cross sections are typically greater than one, our estimates of the sensitivity are conservative.

Note that the sample of signal events we generate include $T$-$t$ fusion processes (see Fig.~\ref{fig:FeynmanDiagram}), as well as $\bar T T$ pair-production, with $T$ decaying to $t\phi'$.  There is little interference between these processes, which can be distinguished because, in the case of $\bar T T$ pair-production, the invariant mass of the $t \phi'$ (or, equivalently $t\gamma \gamma$) system is $m(T)$. The cross section for this latter process depends on the $\mathrm{Br}(T \rightarrow t\phi')$, which depends on details of the model which we do not fix.  For simplicity and to be conservative with our discovery projections,, we set $\mathrm{Br}(T \rightarrow t\phi') = 1\%$, which is reasonable and easily realizable in the parameter space of interest (e.g., see Table I).  We find that these fraction of events arising from $\bar T T$ pair-production is negligible.  

\begin{figure}[tbp]
 \begin{center} 
 \includegraphics[width=0.7\textwidth]{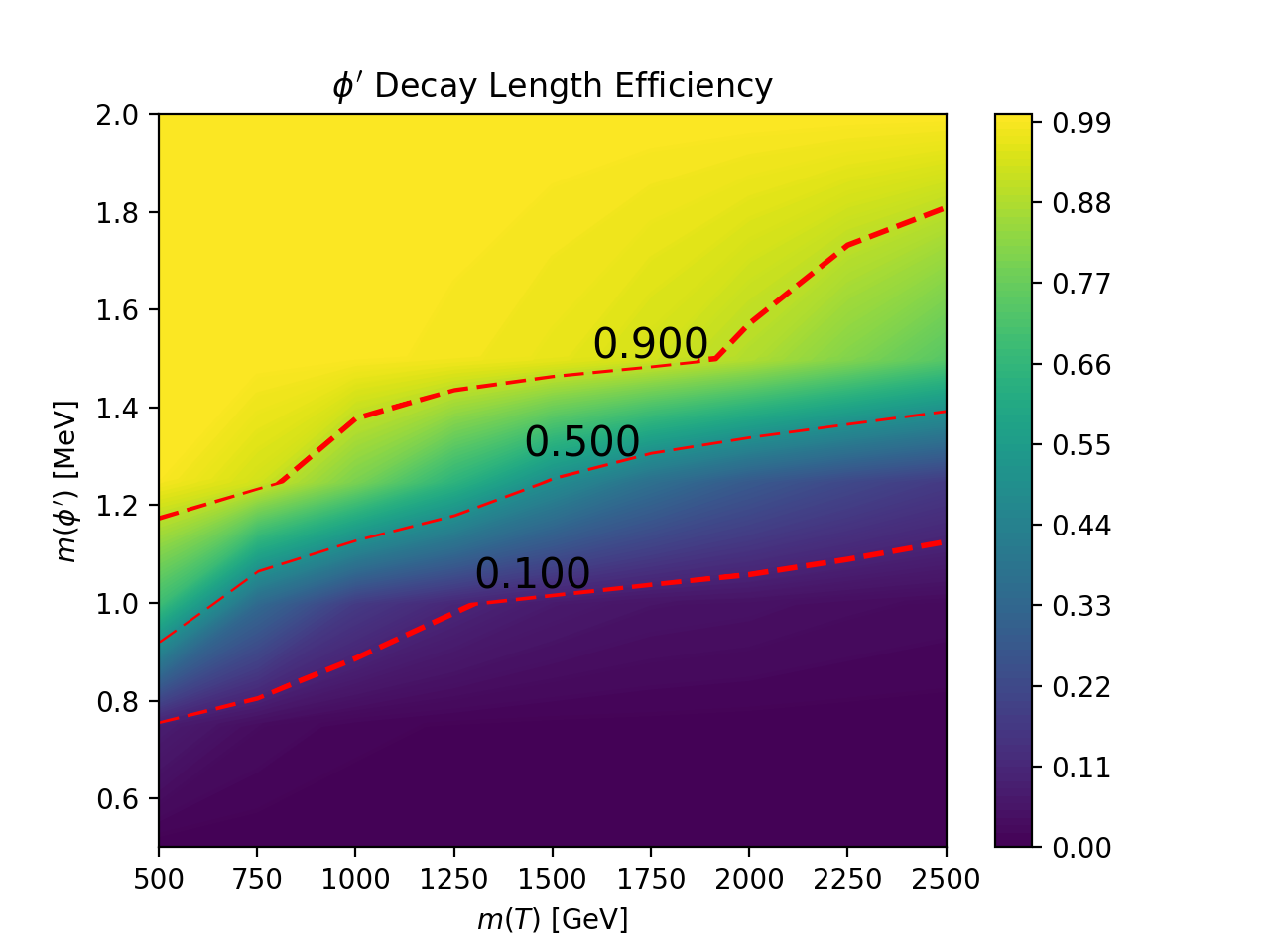}
 \end{center}
 \caption{$\phi'$ decay length efficiency, showing $90\%$, $50\%$, and $10\%$ efficiency levels, assuming $\Gamma_T$ and $\Gamma_{\phi'}$ as described in the text. 
 The contours represent the fraction of events decaying inside the ECAL detector.}
 \label{fig:Phi_Decay_Length_100Contour}
 \end{figure}

The detection efficiency for $\phi' \rightarrow \gamma \gamma$ decays depends on the decay length. The distance the $\phi'$ particle travels before it decays varies depending on the decay width and boost of the $\phi'$. For scenarios with both sufficiently light and sufficiently boosted $\phi'$, a large portion of $\phi'$s begin to decay outside the detector, representing a change in the final state being reconstructed. This effect represents a lower bound of $\phi'$ mass that can be probed by the proposed diphoton final state. To determine the range in $m(\phi^\prime)$ at which the long lifetime becomes important, we compute the $\phi^\prime$ decay length perpendicular to the $pp$ beam axis, which has the form $L_{\phi^\prime,\perp} = \frac{\sqrt{\gamma_{\phi^\prime}^{2}-1}}{\Gamma_{\phi^\prime}}\sin\theta$. In this equation, $\theta$ is the scattering angle relative to the beam axis and $\gamma_{\phi^\prime}$ is the relativistic boost factor. This quantity is calculated per simulated signal event by utilizing the $\phi^\prime$ pseudorapidity distribution and the momentum of the particle in the laboratory frame. We conservatively require the $\phi^\prime$ to decay before the CMS ECAL. Therefore, events with $L_{\phi^\prime,\perp}$ values greater than 1.29 m cannot be used. The $\phi^\prime$ decay length efficiency was computed by simulating large statistics samples of events for given choices of $m(T)$, $m(\phi^\prime)$ and $\Gamma_{T}$, and calculating the fraction of events in which the $\phi^\prime$ decays within the detector. Figure~\ref{fig:Phi_Decay_Length_100Contour} shows these values as a function of both $m(T)$ and $m(\phi')$, before any selection cuts were applied, taking the $\phi'$ lifetime to scale as $t_{\phi'}\sim(10^{-14}\s)(\mathrm{MeV}/m_{\phi'})^3$ (for further discussion of the $\phi'$ lifetime, see the Appendix).As this figure illustrates, the sensitivity will drop rapidly with $m_{\phi'}$ once the $\phi'$ becomes light enough that an appreciable fraction of $\phi'$ decay outside the detector.
 
Table~\ref{tab:lifetime_150} and ~\ref{tab:lifetime_3000} shows the longest lifetime the $\phi'$ can have for which a 5$\sigma$ discovery is still possible for $150~\fb^{-1}$ and $3000~\fb^{-1}$ integrated luminosity respectively. The exact event selection criteria will be defined in the next section. For now, we note that for longer lifetimes, too few $\phi'$s decay inside the detector to achieve such sensitivity.

\begin{figure}[tbp]
 \begin{center} 
 \includegraphics[width=0.4\textwidth]{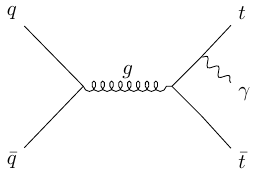}
 \end{center}
 \caption{Typical $t\bar{t}+\gamma$ Feynman diagram, shown with final state radiation (FSR).}
 \label{fig:BackgroundFeynmanDiagram}
 \end{figure}

For this study, various SM backgrounds are considered, including $W$+jets, $Z$+jets, $\gamma$+jets, $\gamma\gamma$+jets, QCD multijet, $t\bar{t}+\gamma$, $t\bar{t}+\gamma\gamma$, and $t\bar{t}+H$ ($H\rightarrow\gamma\gamma$) events. The production of $W$+jets, $Z$+jets, $\gamma$+jets, $\gamma\gamma$+jets, and QCD multijet events are found to be a negligible contribution to the proposed search region due to the low probability of having multiple light quark or gluon jets in each event misidentified as hadronic $t$/$W$ decays, photons, and/or leptons. Additionally, the majority of the QCD multijet processes contain no genuine missing momentum from neutrinos. The SM processes with higgs decays to photons do not contribute because the relatively low boost of the higgs boson results in two clean and well separated photons. It was found that $t\bar{t}+\gamma$ (with the $\gamma$ coming from initial state or final state radiation), as shown in figure~\ref{fig:BackgroundFeynmanDiagram}, represented the dominant irreducible background and $> 99$\% of the total background.

 \section{Event selection criteria}\label{sec:eventselection}
 
\begin{figure}[h]
 \begin{center} 
 \includegraphics[width=0.65\textwidth]{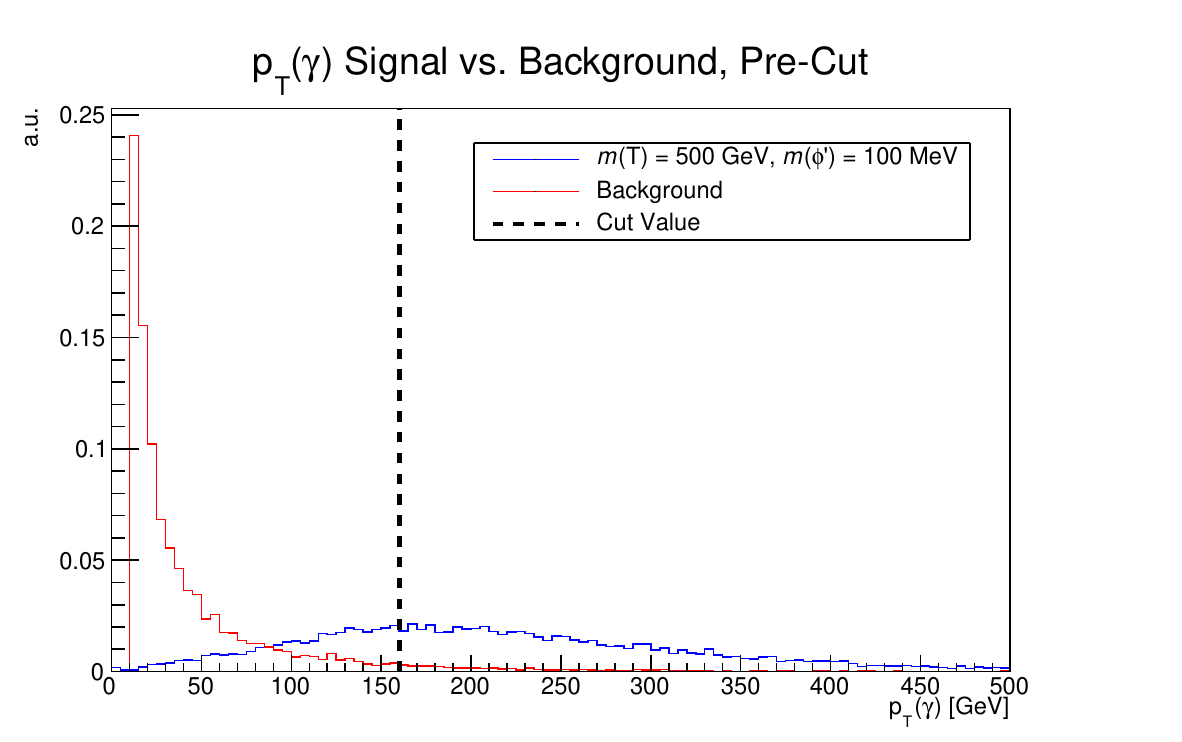}
 \end{center}
 \caption{$p_{T} (\gamma)$ distribution for signal (blue) and background (red) samples.  
 Signal events were generated assuming  $m(T) = 500~\gev$, $m(\phi') = 100~\mev$.  
 The $y$-axis normalization is arbitrary.
 The vertical black dashed line denotes the location of the 
 selection cut.}
 \label{fig:Photon_Pt}
 \end{figure}

\begin{figure}[h]
 \begin{center} 
 \includegraphics[width=0.65\textwidth]{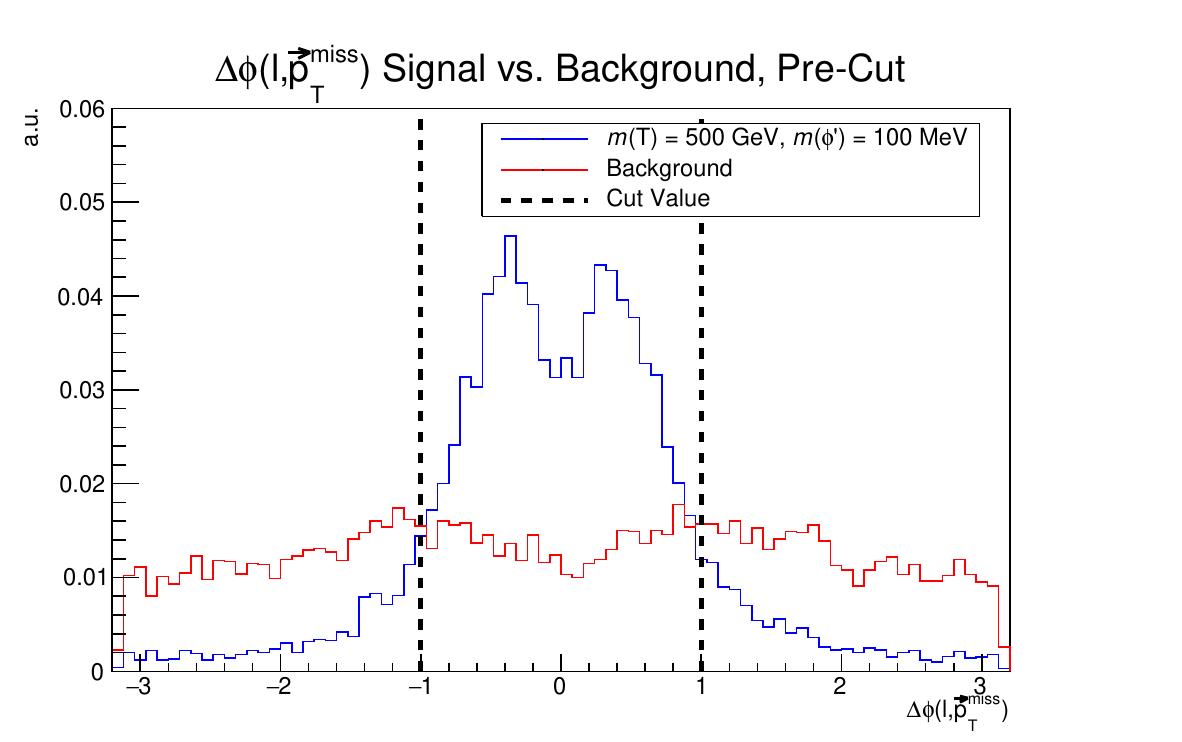}
 \end{center}
 \caption{Distribution of the angle 
between $\ell$ and $\vec{p}_{T}^{\textrm{~miss}}$ ($\Delta\phi (\ell,\vec{p}_{T}^{\textrm{~miss}})$) for 
signal (blue) and background (red) samples.
  Signal events were generated assuming  $m(T) = 500~\gev$, $m(\phi') = 100~\mev$.  
 The $y$-axis normalization is arbitrary.  
 The vertical black dashed lines denotes the location of the 
 selection cut.}
 \label{fig:lepton_MET_delta_phi}
 \end{figure}

\begin{figure}[h]
 \begin{center} 
 \includegraphics[width=0.65\textwidth]{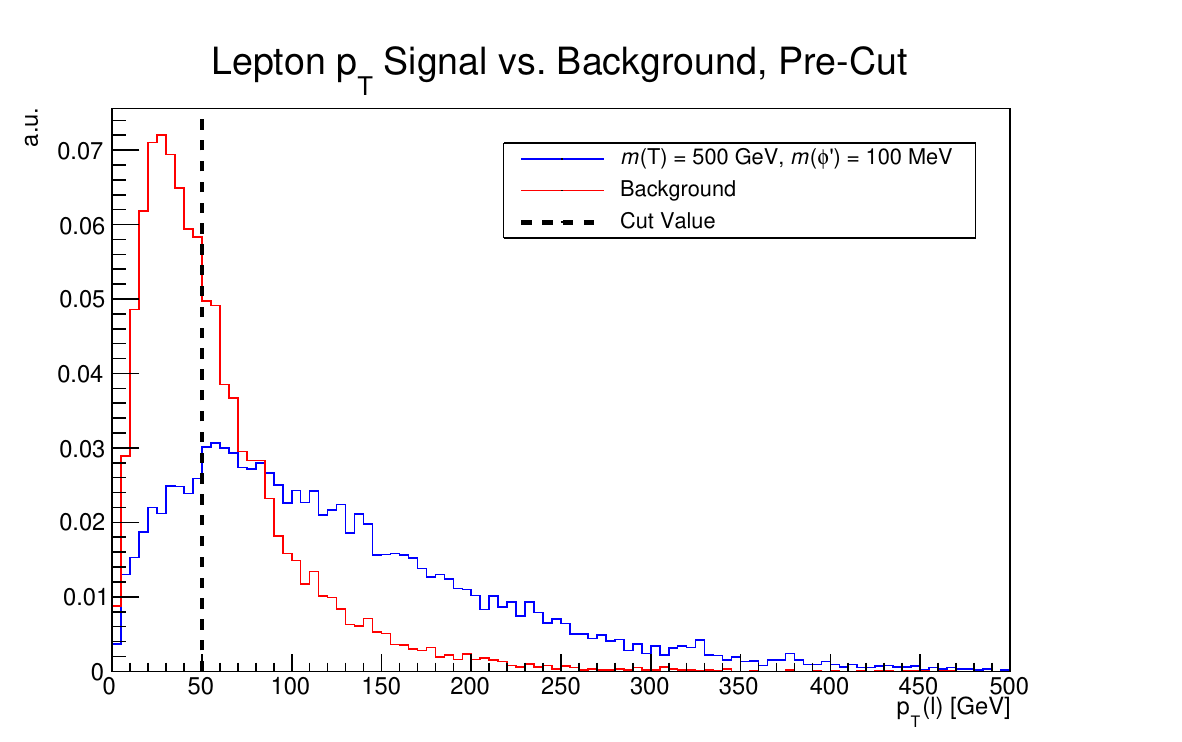}
 \end{center}
 \caption{Distribution of $p_{T}(\ell)$ for signal (blue) and background (red) samples.
 Signal events were generated assuming  $m(T) = 500~\gev$, $m(\phi') = 100~\mev$.  
 The $y$-axis normalization is arbitrary.  
 The vertical black dashed line denotes the location of the 
 selection cut.}
 \label{fig:lepton_pt}
 \end{figure}

 In order to optimize the kinematic selections for maximum discovery potential, a cut-based selection procedure was performed, in which cuts were applied sequentially to optimize signal significance, defined as $N_S/\sqrt{N_S+N_B}$ where $N_S$ and $N_B$ are the number of events in signal and background respectively. We note this particular definition of signal significance is only used for the purpose of optimizing the selections. The final discovery reach is determined with a shape based analysis (described later) using the full reconstructed $T$ mass spectrum. The reconstructed $T$ mass is defined as the transverse mass $m_{T}(b_{\ell},\ell,\vec{p}_{T}^{\textrm{miss}}) = \sqrt{E_{T}(T)^{2} - p_{T}(T)^{2}}$, where $E_{T}(T)$ and $p_{T}(T)$ are the energy and momentum of the resulting Lorentz vector produced by adding together the 4-momentum vectors of the $b$ quark $b_{\ell}$, the lepton, and the missing transverse momentum. For the SM background, the bulk of the $m_{T}(b_{\ell},\ell,\vec{p}_{T}^{\textrm{miss}})$ distribution lies at relatively low reconstructed mass values and decreases exponentially as $m_{T}(b_{\ell},\ell,\vec{p}_{T}^{\textrm{miss}})$ increases. On the other hand, the $m_{T}(b_{\ell},\ell,\vec{p}_{T}^{\textrm{miss}})$ signal distribution appears as a localized bump centered at approximately $m(T)$ (see Results section). For the purpose of the optimization procedure, the values of $N_{S}$ and $N_{B}$ in the significance metric are derived by selecting events near the reconstructed signal $T$ mass peak, $m(T)-2\sigma_{m_{T}}^{exp}(T) < m_{T}(b_{\ell},\ell,\vec{p}_{T}^{\textrm{miss}}) < m(T)+4\sigma_{m_{T}}^{exp}(T)$, where $\sigma_{m_{T}}^{exp}(T)$ is the reconstructed width of the $m_{T}(b_{\ell},\ell,\vec{p}_{T}^{\textrm{miss}})$ distribution, which contains a contribution from the theoretical $T\to bW$ decay width (computed from the model described in the Appendix), but is dominated by experimental/detector resolution. A benchmark scenario of $m(T) =500~\gev$ and $m(\phi')=100~\mev$ was chosen with which to perform the optimization, and the cut values derived therein were applied to all subsequent mass points. The order of the cut optimization was determined by which variables showed the greatest discriminating power between signal and background. Some examples of relevant kinematic distributions are shown in Figs.~\ref{fig:Photon_Pt},~\ref{fig:lepton_MET_delta_phi}, and~\ref{fig:lepton_pt} (all distributions have been normalized to unity). 
 
 Figure~\ref{fig:Photon_Pt} shows the $p_T (\gamma)$ distribution for signal versus background, with the background shape being a falling exponential which is characteristic of initial or final state radiation, while the signal shape has a pronounced hump structure, which is characteristic of decays coming from a boosted object. The signal significance is optimized for $p_{T}(\gamma) > 160$ GeV. Figure~\ref{fig:lepton_MET_delta_phi} shows the $\Delta\phi (\ell,\vec{p}_{T}^{\textrm{~miss}})$ distribution. For signal, the lepton and neutrino which produces the $\vec{p}_{T}^{\textrm{~miss}}$ are the decay products of a very boosted $W$ system (which carries large momentum from the decay of the massive $T$). We therefore expect there to be a small (although non-zero) $\Delta\phi$ between these two very boosted objects. For background however, the semileptonically decaying $t$ carries comparatively less momentum, and therefore its decay products have a distribution with a wider spread of $\Delta\phi(\ell,\vec{p}_{T}^{\textrm{~miss}})$ values. We determined the optimal $\Delta\phi(\ell,\vec{p}_{T}^{\textrm{miss}})$ to be $|\Delta\phi(\ell,\vec{p}_{T}^{\textrm{miss}})| \leq 1$ radians. Figure~\ref{fig:lepton_pt} shows the $p_{T} (\ell)$ distributions  for signal versus background, where the signal lepton carries more momentum due to it being a final decay product of the massive $T$ and therefore peaks at a larger value than the background process. By a similar methodology, we determined the optimal threshold for the $p_{T}$ of the lepton to be $p_{T}(\ell) > 50$ GeV.
 
The identification of leptons and boosted top quarks plays an important role in the signal acceptance and reduction of SM backgrounds, and thus also the projected High-Luminosity LHC (HL-LHC) discovery reach. In the case of the HL-LHC, the reconstruction and identification of leptons and the hadronic decay products of the top quark may be non-trivial due to the presence of a large number of pileup interactions. The importance of pileup mitigation at CMS and ATLAS has been outlined in many papers, such as Ref.~\cite{CMS-PAS-FTR-13-014}. While the expected performance of the upgraded ATLAS and CMS detectors for the HL-LHC is beyond the scope of this work, the studies presented in this paper do attempt to provide reasonable expectations by conservatively assuming some degradation in lepton and hadron identification efficiencies, based on Ref.~\cite{CMS-PAS-FTR-13-014}, and considering the case of 140 average pileup interactions. For muons (electrons) with $|\eta|< 1.5$, the assumed identification efficiency is 90\% (85\%), with a 0.3\% (0.6\%) misidentification rate. The performance degrades linearly with $\eta$ for $1.5 < |\eta| < 2.5$, and we assume an identification efficiency of 65\% (60\%), and a 0.5\% (1.0\%) misidentification rate, at $|\eta| = 2.5$. Similarly, the charged hadron tracking efficiency, which contributes to the jet clustering algorithm and $\vec{p}_{T}^{\textrm{~miss}}$ calculation, is 97\% for $1.5 < |\eta| < 2.5$, and degrades to about 85\% at $|\eta| = 2.5$. These potential inefficiencies due to the presence of secondary $pp$ interactions contribute to how well the lepton and top kinematics can be reconstructed. 

In addition to the $p_{T}(\gamma)$, $|\Delta\phi(\ell,\vec{p}_{T}^{\textnormal{miss}})|$, and $p_{T}(\ell)$ requirements motivated by Figs.~7-9, events with a fully merged top quark system must have a reconstructed FatJet mass compatible with the top quark mass, requiring $m_{\textnormal{reco}}(t) \in [120,220]$. Finally, we impose a modest missing momentum requirement of $p_{T}^{\textnormal{miss}} > 20$ GeV. While the $m_{\textnormal{reco}}(t)$ and $p_{T}^{\textnormal{miss}}$ selections are highly efficient for signal events and $t\bar{t}+\gamma$ background events, they help to ensure the negligible contribution from other SM backgrounds with pairs of vector bosons, $Z/\gamma^{*}\to \ell\ell$ with associated jets, $W\to \ell\nu$ with associated jets, and QCD multijet production of light quarks. Table~\ref{tab:Selection_Cuts} shows the final derived selection cuts and values, with the listing order indicating the order in which the cuts were optimized.

We would like to note that although the results of our optimization procedure are shown for a benchmark signal model with $m(T) = 500$ GeV, the optimized cuts/selections remain similar for the larger $m(T)$ values since the analysis strategy is to look for a localized bump in the high end of the $m_{T}(b_{\ell},\ell,\vec{p}_{T}^{\textrm{miss}})$ spectrum, which is highly correlated to the kinematics of the objects we’re optimizing. Therefore, choosing events with larger (smaller) reconstructed $m_{T}(b_{\ell},\ell,\vec{p}_{T}^{\textrm{miss}})$ effectively means a harder (softer) $p_{T}(b_{\ell})$, $p_{T}(\ell)$, and $\vec{p}_{T}^{\textrm{miss}}$ spectrum. For this reason, although the derived optimal kinematic thresholds shown in Table 3 are similar for all $m(T)$ scenarios, the reconstructed mass window requirement (or the binned likelihood fit of the full $m_{T}(b_{\ell},\ell,\vec{p}_{T}^{\textrm{miss}})$ distribution, discussed in the next section) is taking care of ensuring that the set of events used to assess the presence of signal (i.e. the signal significance calculation) is appropriate for each signal point in the parameter space. Finally, we once again point out that the large mass difference between $T$ and $\phi'$ results in kinematic distributions for $\phi'$ that were similar regardless of $m(\phi')$ mass (for masses up to 10 GeV). Therefore, the results of the optimization procedure do not depend on the mass of $\phi’$.
 
 \begin{table}[tbp]
\centering
\begin{tabular}{c|l}
\hline
Selection Parameter & Value  \\

&\\

\hline 
$p_{T}(\gamma)$ & $\geq 160$ GeV\\
$|\Delta\phi(\ell,\vec{p}_{T}^{\textrm{~miss}})|$ & $\leq$ 1 rads\\
$p_{T}(\ell)$& $\geq 50$ GeV\\
$\eta(\ell)$ & $\leq 2.5$\\
$p_{T}(b_{\ell})$ & $\geq 140$ GeV\\
$\eta(b_{\ell})$ & $\leq 2.4$\\
$|\vec{p}_{T}^{\textrm{~miss}}|$ & $\geq$ 20 GeV \\
$m_{\textrm{reco}}(t)$  & [120,220] GeV\\
$m_T(b_{\ell},\ell,\vec{p}_{T}^{\textrm{~miss}})$ lower bound & $m(T)-2*\sigma_{m_{T}}^{exp}(T)$\\
$m_T(b_{\ell},\ell,\vec{p}_{T}^{\textrm{~miss}})$ upper bound & $m(T)+4*\sigma_{m_{T}}^{exp}(T)$ \\
\hline
\end{tabular}
\captionsetup{justification   = RaggedRight,
             labelfont = bf}
\caption{\label{tab:Selection_Cuts} Selection cut values for all mass points in this analysis.  }
\end{table}

  \begin{figure}[tbp]
 \begin{center} 
 \includegraphics[width=0.65\textwidth]{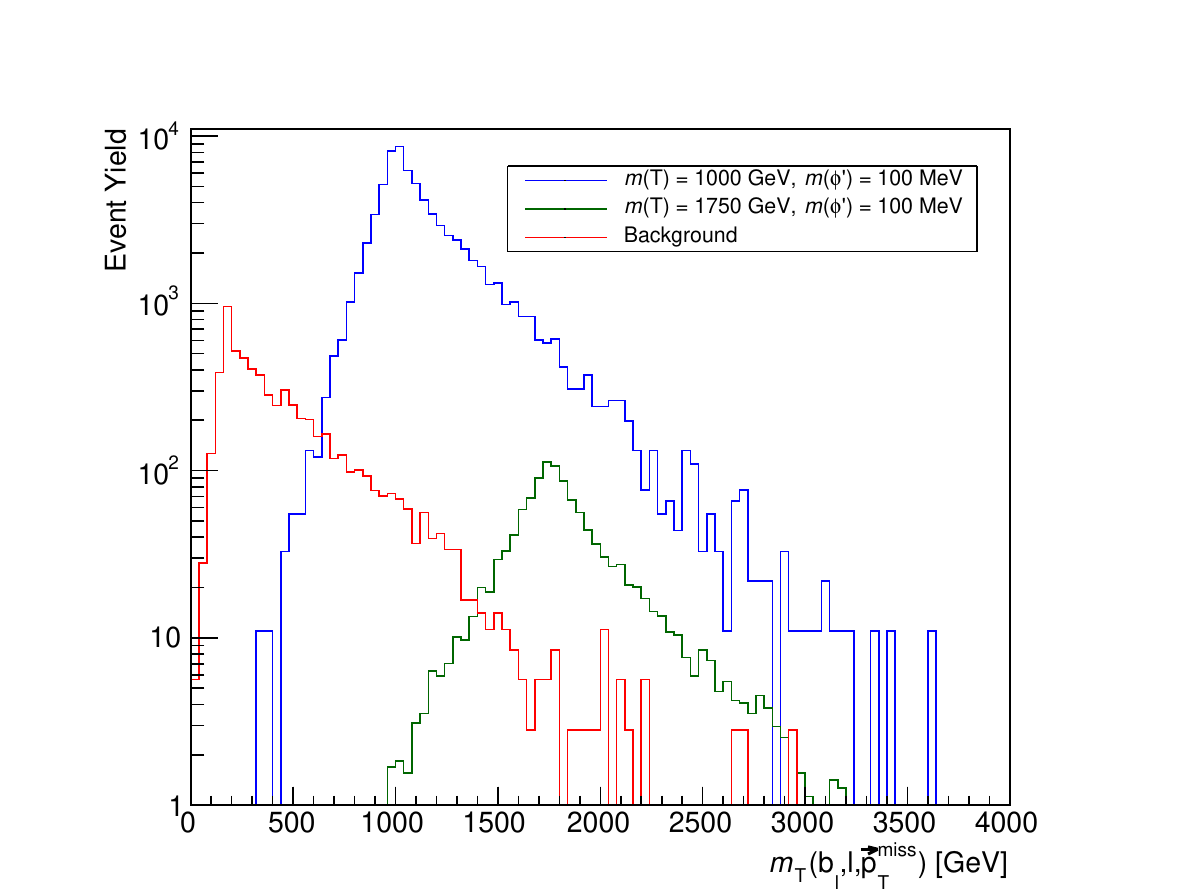}
 \end{center}
 \caption{Distribution of $m_{T}(b_{\ell},\ell,\vec{p}_{T}^{miss})$ for signal and background (red) after applying selection cuts on all parameters except $m_{T}(b_{\ell},\ell,\vec{p}_{T}^{miss})$, assuming $\mathcal{L} = 3000$ fb$^{-1}$. Two $m(T)$ mass points are shown: $m(T) = 1000$ GeV (blue) \& $m(T) = 1750$ GeV (green). In both signal benchmark scenarios, $m(\phi^{\prime})$ is 100 MeV.}
 \label{fig:SR_Plot}
 \end{figure}

 \section{Results}\label{sec:Results}
 
 Figure~\ref{fig:SR_Plot} shows the expected background and signal yields in bins of $m_{T}(b_{\ell},\ell,\vec{p}_{T}^{\textrm{miss}})$. A few signal benchmark points are considered and the yields are normalized to cross section times integrated luminosity of 3000 fb$^{-1}$. The signal cross section depends on a few details of the model, such as the $T-t-\phi'$ coupling.  The scenario we adopt for this analysis is described in the Appendix. The predicted background yield in the $m_{T}(b_{\ell},\ell,\vec{p}_{T}^{\textrm{miss}})$ mass window of [930,1140] GeV is $N_{B} = 250.5 \pm 22.8$, while the signal yield for $m(T)=1000$ GeV within the same mass window is $N_{S} = 2933.3 \pm 52.3$. For the $m_{T}(b_{\ell},\ell,\vec{p}_{T}^{\textrm{miss}})$ mass window of [1627.5,1995] GeV, the predicted background yield is $N_{B} = 27.2 \pm 7.5$, while the signal yield for $m(T)=1750$ GeV is $N_{S} = 61.6 \pm 0.9$. The uncertainties on these numbers represent statistical uncertainties due to the number of raw simulated events passing the selection criteria.
 
 To assess the expected experimental sensitivity of this search at the LHC, we followed a profile binned likelihood test statistic approach, using the expected bin-by-bin yields in the reconstructed $m_{T}(\ell,b,\vec{p}_{T}^{\textrm{~miss}})$ distribution. The signal significance $z$ is determined using the probability of obtaining the same test statistic with the background-only hypothesis and the signal plus background hypothesis, defined as the local p-value. The value of $z$ corresponds to the point where the integral of a Gaussian distribution between $z$ and $\infty$ results in a value equal to the local p-value. The calculation of the signal significance is performed assuming two values for the total integrated luminosity at the LHC: ($i$) 150 fb$^{-1}$, which is approximately the amount of $pp$ data already collected by ATLAS and CMS; and ($ii$) 3000 fb$^{-1}$, expected by the end of the High Luminosity LHC era. The $m_{T}$-based signal significance calculations were performed using the ROOFit \cite{Moneta:2010pm} toolkit developed by CERN.

Various sources of systematic uncertainties are considered in the signal significance calculation. These uncertainties, based on both experimental and theoretical constraints, are incorporated in the test statistic as nuisance parameters, assuming log-normal priors for normalization parameters, and Gaussian priors for uncertainties on the shape of the $m_{T}(\ell,b,\vec{p}_{T}^{\textrm{~miss}})$ distribution. Experimental systematic uncertainties on $\gamma$ identification, and on reconstruction and identification of $b$-jets and top-tagged jets were considered. For $\gamma$ identification, Refs.~\cite{CMS:2016xbb,ATLAS:2014pcp} reports a systematic uncertainty of about 10\%. However, to account for possible effects from the merged photon system, we assume a conservative 15\% uncertainty which is uncorrelated between signal and background processes, and correlated across $m_{T}$ bins for each process. For experimental uncertainties related to the reconstruction and identification of boosted top quarks, a 20\% value was included (independent of $p_{T}$ and $\eta$ of the top-tagged system), following the results from Refs.~\cite{CMS:2020poo,ATLAS:2018wis}. The uncertainty from $b$-jet identification is 10\%, while efficiencies for the electron and muon reconstruction, identification, and isolation requirements have an uncertainty of 2\%. The uncertainties on the reconstruction of $\vec{p}_{T}^{\textrm{~miss}}$ are due to the uncertainties on proper jet energy measurements. We assumed 2-5\% jet energy scale uncertainties, depending on $\eta$ and $p_{T}$, resulting in shape-based uncertainties on $m_{T}$ that range from 3\% to 6\%, depending on the $m_{T}$ bin.

 \begin{figure}[tbp]
 \begin{center} 
 \includegraphics[width=0.65\textwidth]{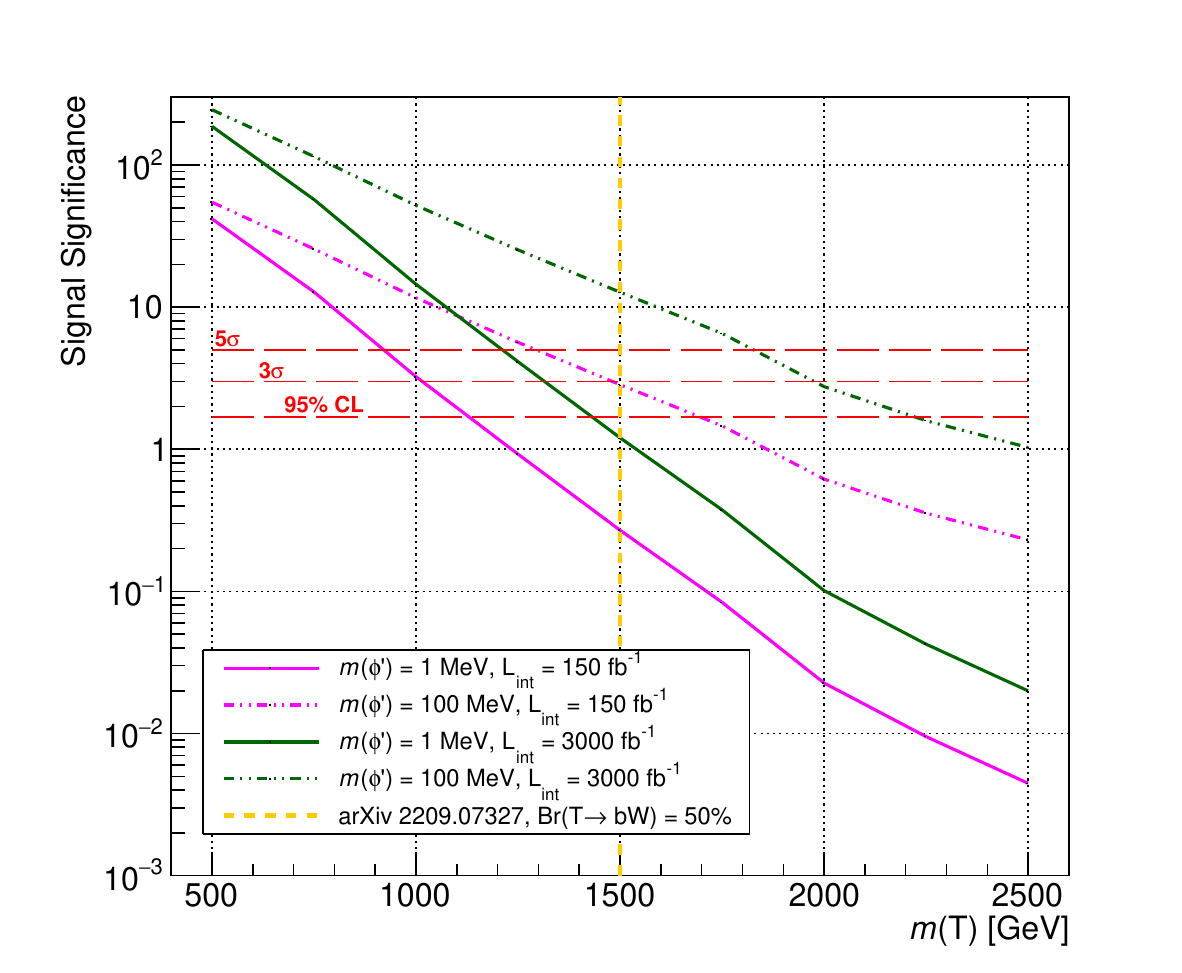}
 \end{center}
 \caption{Signal significance as a function of $m(T)$, for scenarios with $m(\phi')=1$ MeV (magenta) and  $m(\phi')=100$ MeV (green) and for scenarios with $\mathcal{L}_{int}=150\;\mathrm{fb}^{-1}$ (solid lines) and $\mathcal{L}_{int}=3000\;\mathrm{fb}^{-1}$ (dashed lines).   The dashed vertical yellow line indicates the lower limit on $m(T)$ found in~\cite{CMS13TeVVLQ}, 
 assuming $\mathrm{Br}(T \rightarrow bW) = 50\%$.}
 \label{fig:Signal_Significance}
 \end{figure}
 
 \begin{figure}[tbp]
 \begin{center} 
 \includegraphics[width=0.7\textwidth]{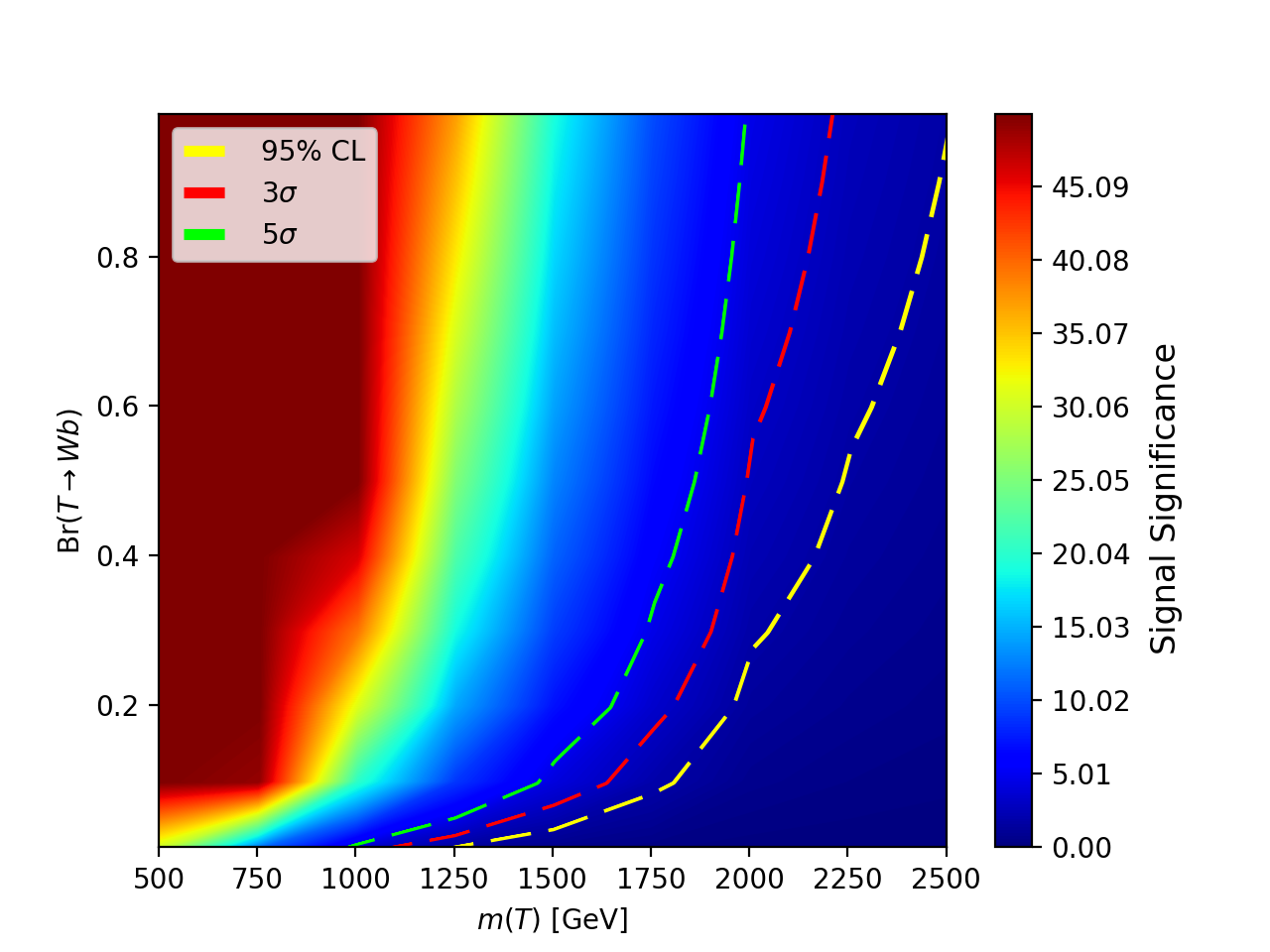}
 \end{center}
 \caption{Signal significance as a function of $m(T)$ and Br($T\rightarrow Wb$), for $m(\phi')$ = 100 MeV. The heat map levels correspond to signal significance, while overlaid dashed lines correspond to the 95\% confidence level (yellow), 3$\sigma$ discovery potential (green), and 5$\sigma$ discovery potential (red), assuming $\mathcal{L}_{int}=3000$ fb$^{-1}$. }
 \label{fig:SignalSig_BF_mT}
 \end{figure}
 
In addition, theoretical uncertainties were included in order to account for the set of parton distribution functions (PDFs) used to produce the simulated  signal and background samples. The PDF uncertainty was calculated following the PDF4LHC prescription~\cite{Butterworth:2015oua}, and results in a 3-10\% systematic uncertainty, depending on the process. The effect of the chosen PDF set on the shape of the $m_{T}$ distribution is negligible. The absence of higher-order contributions to the signal cross sections affect the signal acceptance calculation. This effect is estimated by varying the renormalization and factorization scales by a factor of two with respect to the nominal value, and by considering the full change in the yields. They are found to be small (3.5\%).

Figure~\ref{fig:Signal_Significance} shows the results of the expected signal significance for different  $m(T)$ and $m(\phi')$ scenarios. For the 150 fb$^{-1}$ scenario, it is feasible to exclude (at 95\% confidence level)  $ m(T) < 1.7$ $(1.1)$ TeV for $m(\phi') = 100$ $(1)$ MeV. In order to compare the expected reach of the proposed search with existing searches from CMS and ATLAS, Figure~\ref{fig:Signal_Significance} also shows the current observed upper limit on $m(T)$ from CMS/ATLAS under similar branching fraction considerations. The expected 95\% exclusion of $m(T) < 1.7$ TeV  from the proposed search is to be compared with the $m(T) < 1.4$ (1.5) TeV expected (observed) exclusion from the current LHC searches~\cite{CMS13TeVVLQ}. Similarly, for the 3000 fb$^{-1}$ scenario, the 5$\sigma$ ($3\sigma$) reach includes $m(T) < 1.8$ ($2$) TeV for $m(\phi') = 100$ MeV, while the expected exclusion is $m(T) < 2.2$ TeV. These projections are competitive with the projected high-luminosity LHC reach in Refs.~\cite{Oleksii2014,Liu:2018hum}. For $m(\phi') \lesssim 1~\mev$, a large fraction of the signal events are lost due to an average $\phi'$ decay length that exceeds 1.29 m, thus preventing a photon signature in the electromagnetic calorimeter. Therefore, for 3000 fb$^{-1}$ and $m(\phi') = 1~\mev$, the 5$\sigma$ ($3\sigma$) reach degrades to $m(T) < 1.2$ ($1.3$) TeV. Figure~\ref{fig:SignalSig_BF_mT} further demonstrates how signal significance changes when one now varies the branching fraction of $\mathrm{Br}(T\to Wb)$ for various $m(T)$ (assuming $\mathcal{L}_{int}=3000$ fb$^{-1}$). 

Note that, in this scenario, the search for light scalars in association with heavy quarks can provide better detection prospects than the search either light scalars or heavy quarks individually.

\section{Discussion} \label{sec:discussions}
 
 We have demonstrated that the LHC can discover the visible decays of new sub-GeV particles.  The key to its discovery potential is the production of heavy QCD-coupled particles, which then decay to the light particles with large central boost.  The large boost allows the signal from visible decays to be seen above the lower-energy background.  
 
 There are many laboratory experiments, either current  or planned, which focus on the detection of new low-mass particles.  But the LHC can fill a niche which is not necessarily covered by other strategies.  Most beam experiments focus on the production of light particles through processes with low momentum  transfer, in which the produced new particles are  largely co-linear with the beam.  These experiments  must either deal with the large backgrounds which come  with the beam itself, or are limited to the search  for long-lived particles which decay at displaced  detectors that are protected from beam backgrounds  by shielding.  Indeed, searches for low-mass mediators at many beam experiments (even when assuming the mediator decays to visible particles) assume a long lifetime, and thus search either for energy deposited in a far detector, or energy missed in  a near detector.  Such searches are inapplicable to this scenario, in what the mediator decays promptly. 
 
 The LHC search strategy described here can be used to discover the prompt decay of new light particles, because they are produced from processes with large momentum transfer, and thus are emitted in the central region, away from the beam.  We thus see that detection prospects for low-mass particles are enhanced when it is kinematically possible to access the heavy degrees of freedom which arise in the UV completion of the low-energy model.  Indeed, this scenario, in which the dominant coupling of the light scalar is to the top quark, is an example which would be difficult to directly probe at low energy beam experiments.
 
 We considered the specific example of a new gauge group $U(1)_{T3R}$, in which the dark Higgs is a light (sub-GeV) particle that couples to a heavy new vector-like quark.  This new fermion appears as the seesaw partner of the top quark, and appears as part of the UV completion of the low-energy model.  But this scenario was introduced as a toy-model, and the basic idea will generalize to other models.  
 
We also note that other, related channels can also be used to detect low mass scalars produced in conjunction with  new heavy particles. For example, if the scalar is slightly heavier than the range considered here, decays to $\mu^+ \mu^-$ will be possible.  This channel should also provide good detection prospects.  It would be interesting to study these prospects in a future work.

\acknowledgments

We are grateful to Noah Steinberg and James D.~Wells for useful discussions.  The work of BD, SG and TK is supported in part by DOE grant DE-SC0010813. The work of SG is also supported in part by National Research Foundation of Korea (NRF) Grant No. NRF-2019R1A2C3005009 (SG). The work of JK is supported in part by DOE grant DE-SC0010504. AG and DJ are supported in part by NSF Award PHY-1945366.

\appendix

\section{$U(1)_{T3R}$ and the Universal Seesaw Mechanism} \label{sec:model}

We develop here a concrete model in which we obtain a new heavy fermion $T$ with the same unbroken quantum numbers as the top quark, and a low-mass scalar $\phi'$ which is neutral under the unbroken symmetries of the SM.

We consider a scenario in which a set of right-handed SM up-type and down-type fermions are charged under the gauge group $U(1)_{T3R}$, with charge $\pm Q$, respectively.  Provided that we couple an up-type quark, down-type quark, charged lepton, and right-handed neutrino to $U(1)_{T3R}$, all gauge anomalies will manifestly cancel.  For the purpose of our analysis, we assume that right-handed $t$-quarks couple to $U(1)_{T3R}$.  Since much of the sensitivity of our analysis is driven by the decay products of $t$ and $T$, the identity of the other fermions coupling to $U(1)_{T3R}$ will not affect our analysis.

For simplicity, we will refer to the down-type quark, charged lepton, and right-handed neutrino as $b$, $\ell$, and $\nu_R$ respectively.

We assume that a dark Higgs field $\phi$ also has charge $Q$ under $U(1)_{T3R}$ and is neutral under all other gauge symmetries.  $U(1)_{T3R}$ is spontaneously broken when this field gets a vev $\langle \phi \rangle =V$.  We may write the excitation about this vev as $\phi = V + (1/\sqrt{2}) \phi'$, where $\phi'$ is a real scalar field (the imaginary part of the excitation about the vev is the Goldstone mode, and is absorbed into the longitudinal polarization of the dark photon, which is the gauge boson of $U(1)_{T3R}$).

Each charged fermion which couples to $U(1)_{T3R}$ has a mass which is now protected by both $U(1)_{T3R}$ and $SU(2)_L \times U(1)_Y$, and thus is proportional to the Higgs vev ($v$) and the dark Higgs vev ($V$)~\footnote{The mass terms for the neutrino are more complicated, since the neutrino can also have a Majorana mass term.  For our analysis, we will not need to consider them further.}.  In the effective field theory defined at the electroweak scale, the charged fermion masses thus arise from dimension 5 operators given by
\bea
{\cal L}_{\rm eff} &=& ... - \frac{\lambda_f}{\Lambda_f} 
H^* \phi (\bar f P_R f ) + h.c.,
\eea
with 
\bea
m_f &=& \frac{\lambda_f \langle H \rangle 
\langle \phi \rangle}{\Lambda_f} .
\eea
This non-renormalizable operator  can be 
derived from a UV-complete renormalizable theory if 
we introduce a new vector-like fermion $\tilde \chi_f$, which 
is a singlet under $U(1)_{T3R}$ and has the 
same charges under the SM gauge group 
$SU(3)_C \times SU(2)_L \times U(1)_Y$ as the 
right-handed fermion $f_R$.

In the UV-complete theory, we may write the operators
\bea
{\cal L} &=& ... - \tilde m_{\chi_f} \bar{\tilde{\chi}}_f 
\tilde \chi_f 
- \lambda_{fL} H (\bar{\tilde{\chi}}_f P_L \tilde f_L) 
\nonumber\\
&\,& 
- \lambda_{fR} \phi^* (\bar{\tilde{\chi}}_f P_R \tilde f) 
+h.c., 
\eea
where $\tilde f$ is a flavor eigenstate, and 
$\tilde f_L$ is the $SU(2)_L$ doublet of which 
$\tilde f$ is 
a component.

\subsection{$T$ couplings }

Taking the vevs and couplings to be real, these  Lagrangian terms yield a mass matrix for the flavor eigenstates $\tilde f$ and $\tilde \chi_f$ given by
\bea
M_f &=& \left(
        \begin{array}{cc}
          \tilde m_{\chi_f} & \lambda_{fL} \langle H \rangle \\
          \lambda_{fR} \langle \phi \rangle & 0 \\
        \end{array}
      \right) .
\eea
The lightest mass eigenvalue, $m_f$, is the mass of the SM fermion, while $m_{\chi_f}$ is the heavier mass eigenvalue. 

We are interested in the case where $f=t$.  To be consistent with experimental literature, we will denote the vector-like heavy quark by $\chi_t = T$ (we replace the Lagrangian mass parameter $\tilde{m}_{\chi_t}$ with $\tilde{m}_{T}$).  Since $T$ is charged under $SU(3)_C$, we will assume that it is heavy, in order to avoid current LHC constraints.  If we assume $m_t^2 \ll m(T)^2$,   we then find
\bea
m_t &\sim& \frac{\lambda_{tL} \langle H \rangle  \lambda_{tR} \langle \phi \rangle}{\sqrt{\tilde m_{T}^2 + |\lambda_{tL} \langle H \rangle|^2 +  |\lambda_{tR} \langle \phi \rangle|^2 } },
\nonumber\\
m(T) &\sim& \sqrt{\tilde m_{T}^2 + |\lambda_{tL} \langle H \rangle|^2 +  |\lambda_{tR} \langle \phi \rangle|^2 }.
\label{eqn:seesaw_mass}
\eea
Defining 
\bea
t_{L} &=& \cos \theta_L \tilde t_L + \sin \theta_L 
\tilde T_L ,
\nonumber\\
t_{R} &=& \cos \theta_R \tilde t_R + \sin \theta_R 
\tilde T_{R} ,
\nonumber\\
T_{L} &=& -\sin \theta_L \tilde t_L + \cos \theta_L \tilde T_{L} ,
\nonumber\\
T_{R} &=& -\sin \theta_R \tilde t_R + \cos \theta_R \tilde T_{R} ,
\eea
we find
\bea
\tan \theta_L &= & \frac{1}{\tilde{m}_{T}} \left(
\frac{m_t^2}{\lambda_{tL}\langle H \rangle} 
-\lambda_{tL}\langle H \rangle \right) ,
\nonumber\\
\tan \theta_R &= & \frac{1}{\tilde{m}_{T}} \left(
\frac{m_t^2}{\lambda_{tR}\langle \phi \rangle} 
-\lambda_{tR}\langle \phi \rangle \right) .
\eea

We can parameterize this model in terms of the $\lambda_{tL}$, $\lambda_{tR}$ and $m(T)$ (in terms of which, we can solve for the parameter $\tilde m_T$). For the benchmark scenarios described in the beginning of Section~\ref{sec:Results} and the analysis in Figure~\ref{fig:Signal_Significance}, we 
take the ansatz  $\lambda_{tL,tR}=1$,  enabling us to compute all relevant production cross sections and decay widths as a function of $m(T)$ and 
$m(\phi')$.

As a result of fermion mixing, $T$ can decay to $b W$ or $tZ$, as well $tH$ or $t\phi'$, provided those states are kinematically accessible. Similarly, $T$ can also potentially decay to $t A'$, if that state is kinematically accessible.  However, we leave the signals associated with this state to future work (this scenario has been studied before, see~\cite{Kim:2019oyh}).The branching fraction to all of these final states can be comparable.  The branching fraction to electroweak gauge bosons depends on powers of small mixing angles, but this is compensated by enhancements to the production of the longitudinal polarization.  This is not surprising, as the coupling of $T$ to $H$ and to the longitudinal polarizations of $W$ and $Z$ are related, as a result of the Goldstone Equivalence Theorem.

\begin{figure}[tbp]
 \begin{center} 

  \includegraphics[width=0.65\textwidth]{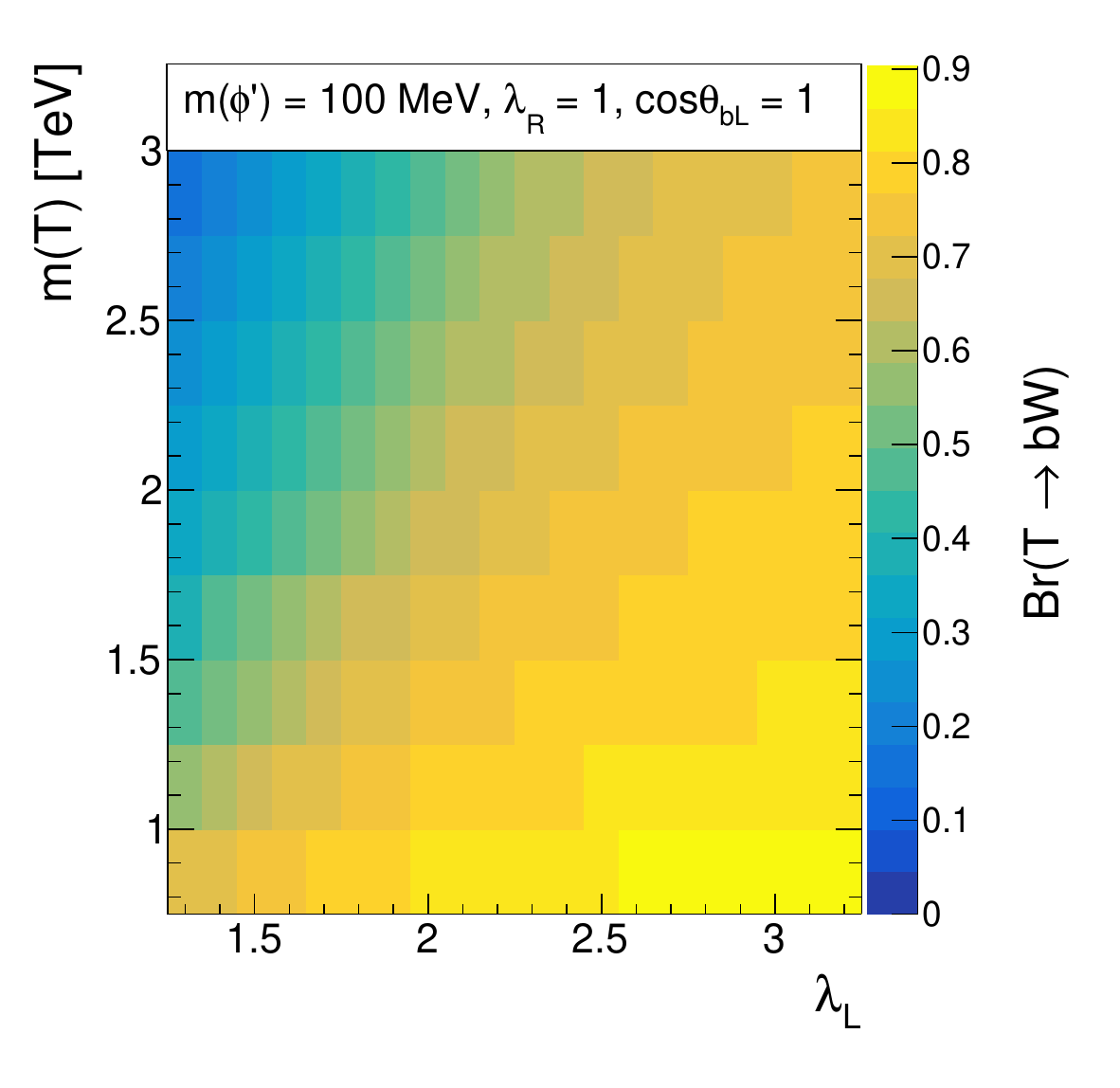}
 \end{center}
 \caption{$T\to bW$ branching fraction as a function of $m(T)$ and $\lambda_{tL}$, for the benchmark scenario with $m(\phi')$ = 100 MeV 
 and $\lambda_{tR}=1$ (assuming no mixing between the $b$-quark and new heavy fermions).}
 \label{fig:BranchingFraction}
 \end{figure}

Figure~\ref{fig:BranchingFraction} shows the calculated $T\to bW$ branching fractions as a function of $m(T)$ and $\lambda_{tL}$, assuming $m(\phi^\prime)$ = 100 MeV and $\lambda_{tR}=1$ (assuming a negligible branching fraction to $tA^\prime$, and assuming no mixing between the $b$-quark and new heavy fermions). We thus see that $\mathrm{Br}(bW) \sim 50\%$ is realizable with appropriate, physically motivated choices of the parameters.   Setting $\lambda_{tL}=1$ also, and $m(T)=2~\tev$ yields the decay widths given in Table~\ref{tab:Width}.

\subsection{$\phi'$ couplings}

The potential for $\phi$ and $H$ can be expressed as 
\bea
V_{\rm pot} &=& -\frac{\mu_\phi^2}{2} \phi \phi^* -\frac{\mu_H^2}{2} H H^*
+ \frac{\lambda}{4} (\phi \phi^*)^2  
\nonumber\\
&\,& + \frac{\lambda_h}{4} (HH^*)^2
+ \frac{\lambda'}{4} (\phi \phi^*) (H H^*) ,
\label{eq:TreeLevelPotential}
\eea
where we will assume $\lambda', \lambda, \lambda_H >0$ and $\mu_{\phi,H}^2 >0$.  In this case, 
we find 
\bea
\langle \phi \rangle  &=& 
\sqrt{(\mu_\phi^2 - \lambda' \langle H \rangle^2/2) 
/ \lambda} ,
\nonumber\\
\langle H \rangle  &=& 
\sqrt{(\mu_H^2 - \lambda' \langle \phi \rangle^2/2) / \lambda_H} ,
\nonumber\\
\tilde m_{\phi'}^2 &=& 2 \lambda \langle \phi \rangle^2 ,
\nonumber\\
\tilde m_{H}^2 &=& 2 \lambda_h \langle H \rangle^2 .
\eea
Provided $\tilde m_{\phi'} < \langle \phi \rangle$, $\tilde m_{H} < \langle H \rangle$, we can 
choose parameters $\mu_{\phi, H}$ and $\lambda, \lambda', 
\lambda_H < 1$ consistent with any choice of 
$\langle \phi \rangle$, $\langle H \rangle$, 
$\tilde m_{\phi',H}^2$

From Eq.~\ref{eqn:seesaw_mass}, we must have $\langle \phi \rangle \gtrsim {\cal O}(\tev)$ if the Yukawa couplings are to be perturbative. Since $m_{\phi'} \lesssim {\cal O}(\gev)$ in our study, the quartic $\lambda$ is necessarily small. Thus, one has to worry that the quartic coupling could run negative at higher energies, resulting in quantum corrections to the potential which could destabilize the vacuum.  To avoid this possibility, we assume that the particle content at masses above $\langle \phi \rangle$ is sufficient to ensure that quartic remain positive.  Since we take $\langle \phi \rangle \gtrsim {\cal O}(\tev)$, there are few experimental constraints on the particle content above this energy.

The mass matrix for the scalars can be written as 
\bea
M_s^2 &=& \left(
        \begin{array}{cc}
          \tilde m_{\phi}^2/2 & \lambda' \langle H \rangle \langle \phi \rangle / 2 \\
          \lambda' \langle H \rangle \langle \phi \rangle / 2 & \tilde m_{H}^2/2 \\
        \end{array}
      \right) .
\eea
Thus, there is a contribution to $\phi'-H$ mixing which is $\sim \lambda' \langle H \rangle \langle \phi \rangle /\tilde m_H^2$.  In addition to the tree-level term,  one expects a one-loop contribution to $\lambda'$ to be generated by diagrams where a fermion runs in the loop.  This contribution will be $\sim (\lambda_{tL}^2 \lambda_{tR}^2 / 16\pi^2) $, yielding a contribution to the mixing angle of magnitude $\sim (\lambda_{tL} \lambda_{tR} /16\pi^2) (m_t m(T)/ \tilde m_H^2)$. In order to avoid constraints on anomalous Higgs decay~\cite{CMS:2022qva,ATLAS:2021gcn,CMS:2021kom}, it is necessary for the mixing angle to be less then ${\cal O}(0.01)$, which is realizable in this scenario.

The dark Higgs $\phi'$ couples to a right-handed up-type quark, down-type quark, charged lepton, and right-handed neutrino. However, because we are interested in the case where $m(\phi')$ is small ($\lesssim {\cal O}(100\mev)$), we will assume that no tree-level decays mediated by renormalizable operators are kinematically allowed.  Instead we will consider the decay process $\phi' \rightarrow \gamma \gamma$, mediated by the higher-dimensional operator $(1 / \Lambda_{\phi'}) \phi' F_{\mu \nu} F^{\mu \nu}$ in the effective field theory defined at low energies.

The coefficient $\Lambda_{\phi'}^{-1}$ would receive contributions from particles heavier than $\phi'$ running in loops, including SM particles as well asother potential new physics couplings to $U(1)_{T3R}$ and to $U(1)_{\rm em}$.  As such, $\Lambda_{\phi'}$ will be model-dependent.   As a benchmark, we can consider the case in which the higher-dimensional operator is generated by integrating out a single heavy fermion, coupled to $\phi'$ and $\gamma$, running in a one-loop diagram.  This diagramwill give a contribution to $\Gamma_{\phi'} / m(\phi')$ which is $\sim {\cal O}(10^{-7}) (m(\phi') / V)^2$ (see, for example, Ref.~\cite{Dutta:2019fxn}).

We see from Table~\ref{tab:lifetime_150} and ~\ref{tab:lifetime_3000}  that, in order for a significant number of $\phi'$ to decay within the detector, one would need $\Gamma_{\phi'}$ to be significantly larger than the benchmark value above.   But if one includes QCD-charge factors and the particle multiplicities of a more complicated dark sector, prompt decays are realizable.

\bibliographystyle{JHEP.bst}
\bibliography{U1T3Rpheno.bib}

\end{document}